\documentclass{article}%
\usepackage{graphicx}
\usepackage{amsmath}
\usepackage{upgreek}
\usepackage{amsfonts}
\usepackage{amssymb}%
\providecommand{\U}[1]{\protect\rule{.1in}{.1in}}

\begin{document}

\author{Antony Valentini\\Augustus College}

\begin{center}
{\LARGE Pilot-wave theory and the search for new physics}

\bigskip

Antony Valentini\footnote{email: a.valentini@imperial.ac.uk}

\textit{Theoretical Physics, Blackett Laboratory, Imperial College London,}

\textit{Prince Consort Road, London SW7 2AZ, United Kingdom.}

\bigskip
\end{center}

We show how pilot-wave theory points to new physics, beyond quantum mechanics,
in three distinct ways. First, generalised cosmological initial conditions,
departing from the Born rule, can lead to observable anomalies in the cosmic
microwave background and in relic cosmological particles. Second, a breakdown
of the Born rule in the deep quantum-gravity regime, with gravitational
corrections that render the Born rule semiclassically unstable, can create
anomalies in Hawking radiation from evaporating black holes. Third, a
regularised equation of motion that remains finite at nodes of the wave
function generates corrections to the Born rule at short distances, while a
natural time-dependent generalisation implies an instability of quantum
equilibrium at short times, effects which may be observable in high-energy collisions.

\bigskip

\bigskip

1 Introduction

2 Generalised initial conditions in cosmology

\qquad{\small 2.1 Quantum relaxation}

\qquad{\small 2.2 Large-scale power deficit in the CMB}

\qquad{\small 2.3 Nonequilibrium relic particles}

3 Quantum gravity and quantum probability

\qquad{\small 3.1 Pilot-wave quantum gravity}

\qquad{\small 3.2 Problem of probability and its solution}

\qquad{\small 3.3 Semiclassical emergence of the Born rule}

\qquad{\small 3.4 Gravitational instability of quantum equilibrium}

4 Regularised pilot-wave theory

\qquad{\small 4.1 New physics at nodes (}${\small \psi=0}${\small )}

\qquad{\small 4.2 Corrections to the Born rule at short distances}

\qquad{\small 4.3 Instability of quantum equilibrium at short times}

\qquad{\small 4.4 Testing the Born rule in high-energy collisions}

5 Conclusion

References

\bigskip

\bigskip

To appear in: \textit{100 Years of Matter Waves}, special issue, Annales de la
Fondation Louis de Broglie.

\bigskip

\section{Introduction}

`\textit{We have to find a new view of the world that has to agree with
everything that is known, but disagree in its predictions somewhere} ... .'
(ref. \cite{Feyn67}, p. 171)

\bigskip

In the search for new physics, it can be useful to have a theory that agrees
with all experiments carried out so far -- and that also suggests new areas
where novel predictions could be found. The pilot-wave approach to quantum
mechanics, pioneered by Louis de Broglie in the 1920s and elaborated by David
Bohm in the 1950s [2--7], provides such a theory. It agrees with quantum
mechanics in a broad regime, but offers three natural points of departure for
new physics: (1) in the early universe, with non-standard initial conditions
that violate the Born rule, (2) in the quantum-gravitational regime, where
standard quantum probabilities are ill-defined, and (3) near nodes of the wave
function, where pilot-wave dynamics breaks down and requires regularisation.
These themes lead to specific predictions. In this paper we outline the basic
motivations and key results.

Pilot-wave theory was first proposed in complete form in 1927 by de Broglie
\cite{deB28,BV09} for a low-energy system of spinless particles whose wave
function $\psi(\mathbf{x}_{1},\mathbf{x}_{2},...,\mathbf{x}_{N},t)$ obeys the
$N$-body Schr\"{o}dinger equation (units $\hbar=1$)%
\begin{equation}
i\frac{\partial\psi}{\partial t}=-\sum_{n=1}^{N}\frac{1}{2m_{n}}\nabla_{n}%
^{2}\psi+V\psi\ . \label{Sch_N}%
\end{equation}
According to de Broglie, the particles move along a definite trajectory
$(\mathbf{x}_{1}(t),...,\mathbf{x}_{N}(t))$ determined by the law of motion%
\begin{equation}
\frac{d\mathbf{x}_{n}}{dt}=\frac{1}{m_{n}}\boldsymbol{\nabla}_{n}S\ ,
\label{dB_N}%
\end{equation}
where $\psi=\left\vert \psi\right\vert e^{iS}$. This implies that, for an
ensemble of systems with the same wave function $\psi$, an arbitrary
distribution $\rho(\mathbf{x}_{1},\mathbf{x}_{2},...,\mathbf{x}_{N},t)$
evolves by the continuity equation%
\begin{equation}
\frac{\partial\rho}{\partial t}+\sum_{n=1}^{N}\boldsymbol{\nabla}_{n}%
\cdot\left(  \rho\mathbf{v}_{n}\right)  =0\ ,
\end{equation}
with a velocity field%
\begin{equation}
\mathbf{v}_{n}=\frac{1}{m_{n}}\boldsymbol{\nabla}_{n}S\ .
\end{equation}
From (\ref{Sch_N}) it follows that $\left\vert \psi\right\vert ^{2}$ obeys the
same continuity equation,%
\begin{equation}
\frac{\partial\left\vert \psi\right\vert ^{2}}{\partial t}+\sum_{n=1}%
^{N}\boldsymbol{\nabla}_{n}\cdot\left(  \left\vert \psi\right\vert
^{2}\mathbf{v}_{n}\right)  =0\ .
\end{equation}
We then have a simple theorem: if $\rho$ and $\left\vert \psi\right\vert ^{2}$
are equal at some initial time, they will remain equal for all time. This is
the state of `quantum equilibrium'%
\begin{equation}
\rho(\mathbf{x}_{1},...,\mathbf{x}_{N},t)=\left\vert \psi(\mathbf{x}%
_{1},...,\mathbf{x}_{N},t)\right\vert ^{2}\ .
\end{equation}
As shown in detail by Bohm in 1952, in this state we obtain agreement with the
usual predictions of quantum mechanics, in particular for quantum
`measurements' of operator observables as usually understood \cite{B52b}.

Similar reasoning applies generally. For a system with configuration $q$ and
wave function $\psi(q,t)$, we have a Schr\"{o}dinger equation%
\begin{equation}
i\frac{\partial\psi}{\partial t}=\hat{H}\psi\ . \label{Sch_q}%
\end{equation}
The system moves along a definite trajectory $q(t)$ determined by the law of
motion%
\begin{equation}
\frac{dq}{dt}=\frac{j(q,t)}{|\psi(q,t)|^{2}}\ , \label{deB_q}%
\end{equation}
where $j=j\left[  \psi\right]  =j(q,t)$ (with a form depending on $\hat{H}$
\cite{SV09}) is the standard quantum current appearing in the
configuration-space continuity equation%
\begin{equation}
\frac{\partial\left\vert \psi\right\vert ^{2}}{\partial t}+\partial_{q}\cdot
j=0 \label{cont_q0}%
\end{equation}
derived from (\ref{Sch_q}). This may be written as%
\begin{equation}
\frac{\partial\left\vert \psi\right\vert ^{2}}{\partial t}+\partial_{q}%
\cdot\left(  |\psi|^{2}v\right)  =0\ ,
\end{equation}
with a velocity field%
\begin{equation}
v(q,t)=\frac{j(q,t)}{|\psi(q,t)|^{2}}\ . \label{v_q}%
\end{equation}
For an ensemble of systems with the same wave function $\psi$, by construction
an arbitrary distribution $\rho(q,t)$ evolves by the same continuity equation%
\begin{equation}
\frac{\partial\rho}{\partial t}+\partial_{q}\cdot\left(  \rho v\right)  =0\ .
\label{cont_q}%
\end{equation}
And so again we have a state of quantum equilibrium%
\begin{equation}
\rho(q,t)=\left\vert \psi(q,t)\right\vert ^{2}\ , \label{Born_q}%
\end{equation}
whose predictions agree with quantum mechanics. This construction applies, for
example, to field theory on a background spacetime (flat or curved), where $q$
is identified with the field configuration on a spacelike slice at global time
$t$.\footnote{Because the dynamics is nonlocal for entangled degrees of
freedom, the dynamics is specified with respect to a preferred slicing of
spacetime with preferred time parameter $t$ \cite{AV08a}.}

Now, as it stands, pilot-wave theory might appear as just another
reformulation of known physics, with nothing measureably new to offer. But in
fact, careful inspection reveals three natural mechanisms for new physics,
which we briefly sketch here, with further details below.

First, the theory provides an obvious generalisation to `quantum
nonequilibrium' ensembles with [10--21]%
\[
\rho(q,t)\neq\left\vert \psi(q,t)\right\vert ^{2}\ .
\]
Here the equations of motion -- (\ref{Sch_q}) for $\psi$ and (\ref{deB_q}) for
$q$ -- remain the same, it is only the initial conditions (over an ensemble)
that differ from usual. There is, of course, a conceptual difference between
immutable laws of motion and variable initial conditions. If we accept
(\ref{Sch_q}) and (\ref{deB_q}) as basic laws, it is clear that in pilot-wave
theory the Born rule (\ref{Born_q}) does not have the same status: the Born
rule is not a law but merely an initial condition.\footnote{For a thorough
discussion see ref. \cite{AV20}.} Therefore, the theory naturally invites us
to consider more general initial conditions with $\rho\neq\left\vert
\psi\right\vert ^{2}$. The potentially observable consequences of this simple
step, in particular for cosmology, are outlined in Section 2.

Second, in the context of quantum gravity, pilot-wave theory suggests that
there is no fundamental equilibrium state at the Planck scale and,
furthermore, that at lower energies small quantum-gravitational corrections
can render the Born rule unstable, enabling initial equilibrium $\rho
=\left\vert \psi\right\vert ^{2}$ to evolve into final nonequilibrium
$\rho\neq\left\vert \psi\right\vert ^{2}$ \cite{AV21,AV23}. While these
effects are expected to be very tiny, they may have observable consequences in
the radiation from exploding primordial black holes (Section 3).

Third, the theory predicts its own demise at nodes where $\psi=0$ and the
velocity field (\ref{v_q}) diverges (for a broad class of Hamiltonians found
in nature). This problem is usually ignored as nodal regions are of (Lebesgue-
or $\left\vert \psi\right\vert ^{2}$-) measure zero. But as a matter of
principle, new physics must set in sufficiently close to nodes, so as to
regularise the theory in some fashion \cite{AV14}. A simple regularisation of
the dynamics implies small corrections to the equilibrium state close to
nodes. We show how the form of the corrections (though not their overall
magnitude) may be predicted. Generalising to a time-dependent regularisation
again enables initial equilibrium $\rho=\left\vert \psi\right\vert ^{2}$ to
evolve into final nonequilibrium $\rho\neq\left\vert \psi\right\vert ^{2}$
\cite{AV14}. Such effects might potentially appear in high-energy collision
experiments, in the form of smeared cross sections, or as anomalous spin or
polarisation probabilities (Section 4).

\section{Generalised initial conditions in cosmology}

The application of quantum mechanics to the early universe is a fairly mature
science, at least for quantum systems on a background classical spacetime. We
know that small temperature anisotropies in the cosmic microwave background
(CMB) were seeded by small inhomogeneities in the early universe. According to
inflationary cosmology, those early inhomogeneities were in turn seeded by
primordial quantum fluctuations at very early times [25--27]. The `primordial
power spectrum' measured by the \textit{Planck} satellite then provides a
record of the Born rule at work in the very early universe. This means that
measurements of the CMB can be employed to test quantum mechanics, in
particular the Born rule, at the earliest moments in the history of our
universe \cite{AV07,AV10}.

\subsection{Quantum relaxation}

According to pilot-wave theory, the Born rule $\rho=\left\vert \psi\right\vert
^{2}$ (applied for example to a field on a background spacetime) is not a law
of nature but merely a state of statistical equilibrium, analogous to thermal
equilibrium in classical physics [10--21]. It can be understood as having
arisen by a past process of dynamical relaxation, whereby an initial
nonequilibrium ensemble with $\rho\neq\left\vert \psi\right\vert ^{2}$ evolves
-- via the continuity equation (\ref{cont_q}) -- towards the equilibrium state
$\rho=\left\vert \psi\right\vert ^{2}$ (on a coarse-grained level). This can
be quantified by a decrease of the coarse-grained $H$-function%
\begin{equation}
\bar{H}(t)=\int dq\ \bar{\rho}\ln(\bar{\rho}/\overline{|\psi|^{2}})
\end{equation}
(for coarse-grained densities $\bar{\rho}$ and $\overline{\left\vert
\psi\right\vert ^{2}}$), which satisfies a coarse-graining $H$-theorem
$\bar{H}(t)\leq\bar{H}(0)$ (assuming no initial fine-grained structure at
$t=0$) \cite{AV91a,AV92,AV01}. Extensive numerical simulations have
illustrated quantum relaxation in a variety of circumstances, with an
approximately exponential decay \cite{VW05,TRV12,ACV14}%
\begin{equation}
\bar{H}(t)\approx\bar{H}_{0}e^{-t/\tau}%
\end{equation}
(for wave functions that are superpositions of multiple energy eigenstates),
where $\bar{H}\rightarrow0$ implies $\bar{\rho}\rightarrow\overline{|\psi
|^{2}}$. We may then understand the Born rule observed today as having arisen
from relaxation processes taking place in the remote past, presumably close to
the big bang.

Such studies have been extended to field theory on a background expanding
space. Consider a massless scalar field $\phi$ on flat expanding space with
metric%
\begin{equation}
d\tau^{2}=dt^{2}-a^{2}\delta_{ij}dx^{i}dx^{j} \label{cosmo metric}%
\end{equation}
and scale factor $a=a(t)$ (conventionally taking $a_{0}=1$ at time $t_{0}$
today). For an unentangled field mode with Fourier component%
\begin{equation}
\phi_{\mathbf{k}}=\frac{\sqrt{V}}{(2\pi)^{3/2}}\left(  q_{\mathbf{k}%
1}+iq_{\mathbf{k}2}\right)
\end{equation}
(where $V$ is a normalisation volume and $q_{\mathbf{k}1}$, $q_{\mathbf{k}2}$
are real), the wave function $\psi_{\mathbf{k}}(q_{\mathbf{k}1},q_{\mathbf{k}%
2},t)$ satisfies the two-dimensional Schr\"{o}dinger equation%
\begin{equation}
i\frac{\partial\psi_{\mathbf{k}}}{\partial t}=-\frac{1}{2a^{3}}\left(
\frac{\partial^{2}}{\partial q_{\mathbf{k}1}^{2}}+\frac{\partial^{2}}{\partial
q_{\mathbf{k}2}^{2}}\right)  \psi_{\mathbf{k}}+\frac{1}{2}ak^{2}\left(
q_{\mathbf{k}1}^{2}+q_{\mathbf{k}2}^{2}\right)  \psi_{\mathbf{k}}\ ,
\end{equation}
with de Broglie velocities%
\begin{equation}
\dot{q}_{\mathbf{k}1}=\frac{1}{a^{3}}\frac{\partial s_{\mathbf{k}}}{\partial
q_{\mathbf{k}1}},\ \ \ \ \dot{q}_{\mathbf{k}2}=\frac{1}{a^{3}}\frac{\partial
s_{\mathbf{k}}}{\partial q_{\mathbf{k}2}} \label{deB_Fourier}%
\end{equation}
(where $\psi_{\mathbf{k}}=\left\vert \psi_{\mathbf{k}}\right\vert
e^{is_{\mathbf{k}}}$) \cite{AV07,AV10}. These match the usual equations for a
two-dimensional harmonic oscillator with time-dependent mass $m=a^{3}$ and
time-dependent angular frequency $\omega=k/a$ (units $\hbar=c=1$). An
arbitrary ensemble distribution $\rho_{\mathbf{k}}(q_{\mathbf{k}%
1},q_{\mathbf{k}2},t)$ evolves by the continuity equation%
\begin{equation}
\frac{\partial\rho_{\mathbf{k}}}{\partial t}+\frac{\partial}{\partial
q_{\mathbf{k}1}}\left(  \rho_{\mathbf{k}}\dot{q}_{\mathbf{k}1}\right)
+\frac{\partial}{\partial q_{\mathbf{k}2}}\left(  \rho_{\mathbf{k}}\dot
{q}_{\mathbf{k}2}\right)  =0\ ,
\end{equation}
which may be integrated forwards in time for given initial conditions.

\subsection{Large-scale power deficit in the CMB}

At short physical wavelengths $\lambda_{\mathrm{phys}}=a\lambda<<H^{-1}$,
where $H^{-1}=a/\dot{a}$ is the Hubble radius, we find the usual rapid and
efficient coarse-grained relaxation $\bar{\rho}_{\mathbf{k}}\rightarrow
\overline{|\psi_{\mathbf{k}}|^{2}}$ indicated above. However, at longer
wavelengths $\lambda_{\mathrm{phys}}\gtrsim H^{-1}$ we find that relaxation is
retarded or suppressed [19, 31--34]. We can then expect quantum noise to be
suppressed at large cosmological scales, as appears to be observed (though not
without controversy, as the data are especially noisy at large scales)
\cite{Planck16}.

The degree of suppression can be quantified by a nonequilibrium mean square%
\begin{equation}
\left\langle \left\vert \phi_{\mathbf{k}}\right\vert ^{2}\right\rangle
=\left\langle \left\vert \phi_{\mathbf{k}}\right\vert ^{2}\right\rangle
_{\mathrm{QT}}\xi(k)\ ,
\end{equation}
where $\left\langle ...\right\rangle _{\mathrm{QT}}$ is the
quantum-theoretical equilibrium value and $\xi(k)<1$ measures the `squeezing'
of the Born rule as a function of $k$. Extensive numerical simulations support
an approximate form (neglecting small oscillations)%
\begin{equation}
\xi(k)=\tan^{-1}(c_{1}\frac{k}{\pi}+c_{2})-\frac{\pi}{2}+c_{3}\ , \label{xi}%
\end{equation}
with an increasing deficit at small $k$, where the unknown parameters $c_{1}$,
$c_{2}$, $c_{3}$ depend on the initial state and on the time interval
\cite{CV15,CV16}. Such large-scale suppression could have taken place during a
(radiation-dominated)\ pre-inflationary era. Assuming for simplicity that the
spectrum is unaffected by the transition to inflation, we may then expect the
usual inflationary power spectrum \cite{LL00}%
\begin{equation}
\mathcal{P}_{\mathcal{R}}^{\mathrm{QT}}(k)=\frac{4\pi k^{3}}{V}\left\langle
|\mathcal{R}_{\mathbf{k}}|^{2}\right\rangle _{\mathrm{QT}}\propto\frac{4\pi
k^{3}}{V}\left\langle \left\vert \phi_{\mathbf{k}}\right\vert ^{2}%
\right\rangle _{\mathrm{QT}}%
\end{equation}
for primordial `curvature perturbations' $\mathcal{R}_{\mathbf{k}}\propto
\phi_{\mathbf{k}}$ to take the corrected form \cite{AV07,AV10}%
\begin{equation}
\mathcal{P}_{\mathcal{R}}(k)=\mathcal{P}_{\mathcal{R}}^{\mathrm{QT}}%
(k)\xi(k)\ , \label{spec_noneq}%
\end{equation}
with a large-scale deficit. This implies corrections to the angular power
spectrum observed in the CMB \cite{AV07,AV10,CV13,CV15}.

The corrected spectrum (\ref{spec_noneq}) has been fit to CMB data with some
success, though at the time of writing there is no clear statistical
preference for a large-scale deficit of the form (\ref{xi}) \cite{VPV19}. To
confirm or rule out cosmological models with quantum relaxation will require
more detailed predictions. Such models imply small oscillations in $\xi(k)$
\cite{CV15}. We may also expect to find statistical anisotropy, with $\xi$
showing some dependence on the direction of the wave vector $\mathbf{k}$ (at
least at large scales) \cite{AV15}. Further comparison with data is a task for
future work.

\subsection{Nonequilibrium relic particles}

According to inflationary cosmology, the matter in our universe was created at
early times by inflaton decay \cite{PU09}. As the early inflationary expansion
comes to an end, the energy of the inflaton field is converted into matter and
radiation, initiating a hot big bang. If the early inflaton field violates the
Born rule, then so will its decay products \cite{AV08b,UV15}. By this
mechanism, the early hot universe could be filled with nonequilibrium
particles that break the Born rule.

We may ask if such particles could survive to the present day and still be in
nonequilibrium. To avoid relaxation, the particles would need to decouple at
very early times so as to stream freely with negligible
interaction.\footnote{The particles would also need to have effective wave
functions with very few energy modes superposed \cite{UV15}.} Prominent
theories of high-energy physics suggest that dark matter is largely composed
of relic particles that decoupled very early -- for example, gravitinos. It
then seems conceivable that, even today, dark-matter particles could violate
the Born rule \cite{AV01,AV07,AV08b,UV15}.

Today it is not possible to experiment with dark-matter particles directly,
since they interact extremely weakly with ordinary matter. However, we may be
able to observe their annihilation or decay products. For example, pairs of
gravitinos are expected to annihilate into pairs of gamma-ray photons. In
other models, dark-matter particles can decay to produce X-rays. If the parent
particles violate the Born rule, so will the outgoing photons. Among other
phenomena, the resulting spectral lines could have anomalous profiles
\cite{UV20}.

We may then envisage experiments testing the Born rule for X-rays or gamma
rays produced by disintegrating dark matter. The \textit{Fermi Gamma-ray Space
Telescope} has detected a puzzling excess of gamma rays from the Galactic
centre \cite{Fermi17}, which on some models could originate from the
annihilation or decay of dark-matter particles \cite{DM20,DM21}. We then have
a case for testing the Galactic excess gamma-ray photons for deviations from
the Born rule. The incoming photons could impinge on an appropriate two-slit
screen, or perhaps a diffracting crystal or grating, where the resulting
interference or diffraction pattern may be monitored for anomalies. The
incoming photons could also be tested for anomalies in the angular dependence
of their differential scattering cross sections in high-energy collisions or,
more simply and cleanly, the photons could be tested directly for anomalies in
their polarisation probabilities (cf. Section 4.4) \cite{AV07,AV04a}. Such
tests would have to be done aboard a dedicated satellite in space, to avoid
absorption of the incoming gamma rays by the earth's atmosphere. The
forthcoming QUICK%
${{}^3}$
satellite mission includes an interferometer designed to test the Born rule in
space \cite{Ahm24}. Once such a system is fully operational, it could be
deployed to test Galactic excess gamma rays -- or photons from some other
exotic source -- for deviations from the Born rule.

\section{Quantum gravity and quantum probability}

So far we have considered quantum systems on a background classical spacetime.
In quantum gravity, spacetime itself is quantised. This turns out to have
profound implications for the Born rule.

\subsection{Pilot-wave quantum gravity}

In the usual `canonical' approach to quantum gravity, spacetime is foliated by
spacelike slices with 3-metric $g_{ij}$, which is then subject to
quantisation. In the presence of a matter field $\phi$ we obtain the
Wheeler-DeWitt equation ($\hbar=c=16\pi G=1$) \cite{DeW67,Kief12}%
\begin{equation}
\left(  -G_{ijkl}\frac{\delta^{2}}{\delta g_{ij}\delta g_{kl}}-\sqrt
{g}R+\mathcal{\hat{H}}_{\phi}\right)  \Psi=0 \label{W-D}%
\end{equation}
for the wave functional $\Psi=\Psi\lbrack g_{ij},\phi]$, where%
\begin{equation}
G_{ijkl}={\frac{1}{2}}g^{-1/2}(g_{ik}g_{jl}+g_{il}g_{jk}-g_{ij}g_{kl})
\end{equation}
and $\mathcal{\hat{H}}_{\phi}$ is an appropriate contribution from $\phi$.
There is also a constraint%
\begin{equation}
-2D_{j}\frac{\delta\Psi}{\delta g_{ij}}+\partial^{i}\phi\frac{\delta\Psi
}{\delta\phi}=0 \label{cons}%
\end{equation}
(where $D_{j}$ is a spatial covariant derivative) ensuring that $\Psi$ is a
function of the coordinate-independent 3-geometry.

It is noteworthy that $\Psi$ has no explicit dependence on time $t$ (where at
the classical level $t$ labels the foliation by spacelike slices). In the
standard quantum-mechanical formulation, this presents numerous difficulties
which are generally known as the `problem of time' [47--53].

In the pilot-wave formulation, in addition to equations (\ref{W-D}) and
(\ref{cons}) for $\Psi$, we also have de Broglie guidance equations
\cite{Hor94}%
\begin{equation}
\frac{\partial g_{ij}}{\partial t}=2NG_{ijkl}\frac{\delta S}{\delta g_{kl}%
}\ ,\ \ \ \ \frac{\partial\phi}{\partial t}=\frac{N}{\sqrt{g}}\frac{\delta
S}{\delta\phi} \label{deB_QG}%
\end{equation}
for evolving fields $g_{ij}$ and $\phi$, where as usual $\Psi=\left\vert
\Psi\right\vert e^{iS}$ and $N$ is the `lapse function' associated with the
foliation.\footnote{For simplicity and without loss of generality we take the
`shift vector' $N^{i}=0$.} Even though $\Psi$ is static, there are
time-dependent trajectories for $g_{ij}$ and $\phi$.

\subsection{Problem of probability and its solution}

Pilot-wave theory has been extensively applied to the simplified version of
quantum gravity known as quantum cosmology \cite{Vink92}. These applications
include bouncing cosmological models and demonstrations of singularity
avoidance [56--58]. These works discuss and calculate de Broglie-Bohm
trajectories, for example for the cosmological scale factor $a(t)$, without
however discussing probability. There is a technical reason for this:
solutions $\Psi$ of the Wheeler-DeWitt equation are non-normalisable and
cannot define a Born-rule probability density $\left\vert \Psi\right\vert
^{2}$. This is because the Wheeler-DeWitt equation for $\Psi$ on configuration
space is mathematically analogous to a Klein-Gordon equation for a field
$\phi$ on spacetime: the integral $\int dq\ \left\vert \Psi\right\vert ^{2}$
diverges as does the integral $\int d^{3}x\int dt\ \left\vert \phi\right\vert
^{2}$ \cite{Kuch92,Ish93}. In pilot-wave theory, this problem can be solved by
abandoning the Born rule at the fundamental level \cite{AV21,AV23}.

We can illustrate the problem, and its solution, with a simple model of
quantum cosmology. We consider a flat expanding universe with metric
(\ref{cosmo metric}). The universe contains a homogenous matter field $\phi$
with potential $\mathcal{V}(\phi)$. We then have a Wheeler-DeWitt equation
\cite{BKK16}%

\begin{equation}
\frac{1}{2m_{\mathrm{P}}^{2}}\frac{1}{a}\frac{\partial}{\partial a}\left(
a\frac{\partial\Psi}{\partial a}\right)  -\frac{1}{2a^{2}}\frac{\partial
^{2}\Psi}{\partial\phi^{2}}+a^{4}\mathcal{V}\Psi=0 \label{WD_mini}%
\end{equation}
for $\Psi(a,\phi)$, where $m_{\mathrm{P}}^{2}=3/4\pi G$ is the (rescaled)
Planck mass squared. In terms of $\alpha=\ln a$ this can be rewritten as a
two-dimensional Klein-Gordon equation%
\begin{equation}
\frac{1}{m_{\mathrm{P}}^{2}}\frac{\partial^{2}\Psi}{\partial\alpha^{2}}%
-\frac{\partial^{2}\Psi}{\partial\phi^{2}}+2e^{6\alpha}\mathcal{V}\Psi=0
\label{WD_mini_alpha}%
\end{equation}
with a potential term. The free part has the general solution%
\begin{equation}
\Psi=f(\phi-m_{\mathrm{P}}\alpha)+g(\phi+m_{\mathrm{P}}\alpha)\ ,
\label{Psi_free}%
\end{equation}
where $f$ and $g$ are packets travelling in the two-dimensional `spacetime'
$(\alpha,\phi)$. Clearly, the integral $\int\int d\alpha d\phi\ \left\vert
\Psi\right\vert ^{2}$ necessarily diverges. This problem can be solved by
denying that $\left\vert \Psi\right\vert ^{2}$ is a measure of probability. In
pilot-wave theory, in addition to the Wheeler-DeWitt equation (\ref{WD_mini}),
we also have the de Broglie guidance equations \cite{AV21,AV23}%
\begin{equation}
\dot{a}=-\frac{1}{m_{\mathrm{P}}^{2}}\frac{1}{a}\frac{\partial S}{\partial
a}\ ,\ \ \dot{\phi}=\frac{1}{a^{3}}\frac{\partial S}{\partial\phi}\ .
\label{deB_cosmo}%
\end{equation}
For a theoretical ensemble with wave function $\Psi(a,\phi)$, we can consider
an arbitrary (by definition normalisable) probability density $P(a,\phi,t)$
whose time evolution follows the continuity equation%
\begin{equation}
\frac{\partial P}{\partial t}+\frac{\partial}{\partial a}\left(  P\dot
{a}\right)  +\frac{\partial}{\partial\phi}\left(  P\dot{\phi}\right)  =0
\end{equation}
with the velocity field (\ref{deB_cosmo}). This can be integrated forwards in
time, given an arbitrary initial condition $P(a,\phi,t_{0})$ at $t=t_{0}$. By
construction, at all times $\int\int dad\phi\ P=1$ whereas $\int\int
dad\phi\ \left\vert \Psi\right\vert ^{2}=\infty$. Thus, necessarily, the
system is in a perpetual state of quantum nonequilibrium\footnote{While this
does not materially affect the discussion, in fact with our choice of
configuration space the natural (non-normalisable) quantum density is equal to
$a^{2}\left\vert \Psi\right\vert ^{2}$ and not $\left\vert \Psi\right\vert
^{2}$ \cite{AV21,AV23}. Strictly speaking we then have a perpetual
nonequilibrium $P\neq a^{2}\left\vert \Psi\right\vert ^{2}$.}%
\begin{equation}
P\neq\left\vert \Psi\right\vert ^{2}\ .
\end{equation}
If we accept that there is no equilibrium state, and no Born rule, in the deep
quantum-gravity regime, we can consistently discuss ensembles with a
normalised probability density $P$.

We might ask what happens to $P$ as it evolves in time. Clearly, it cannot
relax towards $\left\vert \Psi\right\vert ^{2}$ as occurs in conventional
systems (on a coarse-grained level), since $P$ remains normalised. It has been
suggested that, nevertheless, $P$ might relax towards $\left\vert
\Psi\right\vert ^{2}$ in a local region of configuration space \cite{Sen24}.
However, numerical simulations show that relaxation does not occur even
locally \cite{KV24}.

\subsection{Semiclassical emergence of the Born rule}

At this point it is natural to ask how the theory can account for the observed
Born rule in the semiclassical regime -- that is, for a quantum system on a
background classical spacetime, with an effective time-dependent
Schr\"{o}dinger equation%
\begin{equation}
i\frac{\partial\psi}{\partial t}=\hat{H}\psi\label{Sch_eff}%
\end{equation}
for a standard (normalisable) wave function $\psi(q,t)$.

As usually understood this regime emerges, for example for a quantum field
$\phi$, when the Wheeler-DeWitt wave functional takes the form \cite{KS91}%
\begin{equation}
\Psi\lbrack g_{ij},\phi]\approx\Psi_{\mathrm{WKB}}[g_{ij}]\psi\lbrack
\phi,g_{ij}]\ ,
\end{equation}
where $\Psi_{\mathrm{WKB}}[g_{ij}]$ is a WKB state for the background. The
effective time-dependent wave functional $\psi\lbrack\phi,t]$ arises simply by
evaluating $\psi\lbrack\phi,g_{ij}(t)]$ along a classical background
trajectory $g_{ij}(t)$. We then obtain the usual, much-studied regime, where
we expect to find (coarse-grained)\ relaxation $\rho\lbrack\phi,t]\rightarrow
\left\vert \psi\lbrack\phi,t]\right\vert ^{2}$ to quantum equilibrium.

Thus, despite the absence of a fundamental Born-rule state, nevertheless the
Born rule emerges as required in the semiclassical regime, and so the theory
can still agree with observation.

\subsection{Gravitational instability of quantum equilibrium}

In the semiclassical approximation, the effective Hamiltonian $\hat{H}$ has
tiny quantum-gravitational corrections which have been evaluated
\cite{BKK16,KS91}. Remarkably, there are both Hermitian and non-Hermitian
terms. The non-Hermitian terms are usually discarded by hand, since in
standard quantum mechanics they generate a small violation of probability
conservation. However, we can make sense of those terms in pilot-wave theory:
in their presence, probability is still conserved but the Born rule becomes
unstable \cite{AV21,AV23}. To see this note first that, if we evaluate the de
Broglie velocity for $\phi$ to the same approximation, we find that it takes
the usual form $\dot{\phi}=(N/\sqrt{g})(\delta S/\delta\phi)$ but with $S$ now
equal to the phase of $\psi$. We then have the usual continuity equation%
\begin{equation}
\frac{\partial\rho}{\partial t}+\int d^{3}x\;\frac{\delta}{\delta\phi}\left(
\rho\dot{\phi}\right)  =0
\end{equation}
for $\rho\lbrack\phi,t]$. As for $\left\vert \psi\lbrack\phi,t]\right\vert
^{2}$, writing $\hat{H}=\hat{H}_{1}+i\hat{H}_{2}$ (with $\hat{H}_{1}$,
$\hat{H}_{2}$ Hermitian), we find that the Schr\"{o}dinger equation
(\ref{Sch_eff}) implies a continuity equation%
\begin{equation}
\frac{\partial\left\vert \psi\right\vert ^{2}}{\partial t}+\int d^{3}%
x\;\frac{\delta}{\delta\phi}\left(  \left\vert \psi\right\vert ^{2}\dot{\phi
}\right)  =s\ ,
\end{equation}
with a source term%
\begin{equation}
s=2\operatorname{Re}\left(  \psi^{\ast}\hat{H}_{2}\psi\right)  \ .
\end{equation}
Because the continuity equations for $\rho$ and $\left\vert \psi\right\vert
^{2}$ no longer match, an initial distribution $\rho=\left\vert \psi
\right\vert ^{2}$ can evolve into a final distribution $\rho\neq\left\vert
\psi\right\vert ^{2}$. Small quantum-gravitational corrections can generate
nonequilibrium from initial equilibrium. The timescale for this process has
been estimated:\footnote{From the time derivative of the $H$-function
$H(t)=\int D\phi\ \rho\ln(\rho/\left\vert \psi\right\vert ^{2})$
\cite{AV21,AV23}.}%
\begin{equation}
\tau_{\mathrm{noneq}}\approx\frac{1}{2\left\vert \left\langle \hat{H}%
_{2}\right\rangle \right\vert }\ .
\end{equation}

In the presence of an evaporating Schwarzchild black hole of mass $M(t)$, the
Hamiltonian $\hat{H}_{\mathbf{k}}$ of a scalar field mode $\mathbf{k}$ has a
non-Hermitian correction $i\hat{H}_{2}$ where \cite{KMS94}%
\begin{equation}
\hat{H}_{2}\simeq-\frac{1}{12}\kappa\left(  \frac{m_{\mathrm{P}}}{M}\right)
^{4}\hat{H}_{\mathbf{k}}\ ,
\end{equation}
with $\kappa$ a numerical factor and $m_{\mathrm{P}}=\sqrt{\hbar c/G}%
\simeq10^{-5}\ \mathrm{g}$ (the standard Planck mass). If we take
$\left\langle \hat{H}_{\mathbf{k}}\right\rangle $ to be of order the Hawking
temperature%
\[
k_{\mathrm{B}}T_{\mathrm{H}}=\frac{\hbar c^{3}}{G}\frac{1}{8\pi M}=\frac
{1}{8\pi}m_{\mathrm{P}}c^{2}\frac{m_{\mathrm{P}}}{M}\ ,
\]
we find an estimated timescale \cite{AV21,AV23}%
\begin{equation}
\tau_{\mathrm{noneq}}\sim\frac{48\pi}{\kappa}t_{\mathrm{P}}\left(  \frac
{M}{m_{\mathrm{P}}}\right)  ^{5}\ ,
\end{equation}
where $t_{\mathrm{P}}$ is the usual Planck time. The effect is significant
only in the final stages of evaporation when $M$ is close to $m_{\mathrm{P}}$.
We may then expect the very final burst of Hawking radiation to show
deviations from the Born rule.\footnote{Related effects could also play a role
in resolving the information-loss puzzle \cite{AV07,AV04b,KV20}.}

In practice, we may reasonably hope to observe $\gamma$-ray Hawking radiation
from exploding primordial black holes \cite{Ack18}. If such $\gamma$-rays were
discovered, they could be tested for possible violations of the Born rule,
such as anomalous polarisation probabilities \cite{AV07,AV04a}. Again, such
tests would have to be done aboard a dedicated satellite in space.

\section{Regularised pilot-wave theory}

The effects discussed so far require delicate measurements of radiation from
the early universe, from disintegrating relic particles, or from exploding
primordial black holes. One might ask if there are any worthwhile experiments
that might be conducted in the laboratory, under controlled conditions, and
for which there may be plausible prospects of observing deviations from
quantum mechanics.

As it stands, pilot-wave theory tells us that we are trapped in a state of
quantum equilibrium -- a state of `quantum death' broadly analogous to the
state of `heat death' predicted by classical statistical thermodynamics
[10--21]. As we saw in Section 2, we might find evidence for an early state of
quantum nonequilibrium from careful measurements of the CMB, and residual
nonequilibrium could in principle survive to the present day in certains kinds
of relic cosmological particles. But once the equilibrium state is reached,
there appears to be no realistic means of escape, except for the tiny
quantum-gravitational effects which we described in Section 3 -- and which
seem beyond any hope of laboratory detection.\footnote{We may discount as
completely unrealistic attempts to escape from quantum death simply by waiting
for rare fluctuations away from equilibrium \cite{AV92}.} This somewhat
discouraging conclusion is modified, however, if we carefully re-examine the
fundamental equations of pilot-wave dynamics.

\subsection{New physics at nodes ($\psi=0$)}

The Hamiltonians found in nature are often quadratic in the momenta. The de
Broglie velocity (\ref{v_q}) is then proportional to a phase gradient%
\begin{equation}
\partial_{q}S=\operatorname{Im}\frac{\partial_{q}\psi}{\psi}\ ,
\label{ph_grad}%
\end{equation}
which generally diverges at nodes where $\psi=0$. This problem is usually
ignored, perhaps because nodal regions are of measure zero and the
trajectories tend to avoid them. But physically speaking, a diverging velocity
signals a breakdown of pilot-wave dynamics at $\psi=0$ -- just as a diverging
acceleration signals a breakdown of Newtonian gravity (for a point mass) at
$r=0$. In our view, there must be new physics close to points where $\psi=0$
-- just as in Newtonian gravity there is new physics close to $r=0$ where
general relativity becomes important \cite{AV14}.

It should be emphasised that (single-component) wave functions generically do
have nodes: in $n$ dimensions the simultaneous equations $\operatorname{Re}%
\psi=0$, $\operatorname{Im}\psi=0$ generally have solutions yielding nodal
regions of dimension $n-2$. Such points are often referred to as `phase
singularities', where the phase gradient (\ref{ph_grad}) diverges. They might
also be compared with the spacetime singularities of classical general
relativity (for example at the centre of a black hole). Unlike their
gravitational analogues, however, phase singularities do not attract or absorb
trajectories. Instead, trajectories tend to avoid them or circle around them
-- as we might expect since nodal regions have zero equilibrium probability
(that is, zero $\left\vert \psi\right\vert ^{2}$ measure). We can then
understand why phase singularities are usually disregarded as unproblematic in
applications of de Broglie-Bohm theory. Nevertheless, in principle, new
physics must set in sufficiently close to such points.

Consider the simple example of a hydrogen-like atom in a stationary state%
\begin{equation}
\psi(r,\theta,\phi,t)=\psi_{nlm}(r,\theta,\phi)e^{-iE_{n}t} \label{stat}%
\end{equation}
of energy $E_{n}$ and with the usual quantum numbers $nlm$ (in spherical
coordinates). For $m\neq0$ there is a nodal line $\psi_{nlm}=0$ along the
$z$-axis (where $\sin\theta=0$). We have a deBroglie velocity%
\[
\mathbf{v}=\frac{1}{m_{0}}\boldsymbol{\nabla}S
\]
(denoting the particle mass by $m_{0}$), where%
\begin{equation}
S=m\phi-E_{n}t\ .
\end{equation}
The trajectories circle around the $z$-axis at a fixed distance $d=r\sin
\theta$. The speed (ref. \cite{Holl93}, p. 150)%
\begin{equation}
\left\vert \mathbf{v}\right\vert =\frac{m}{m_{0}}\frac{1}{d}%
\end{equation}
diverges as $d\rightarrow0$.

We might ask if this problem could be merely an artifact of the low-energy
theory. Perhaps such divergences are absent in the pilot-wave theory of
high-energy physics.\footnote{For a recent review see ref. \cite{AV24}.}
Support for this appears to come from the pilot-wave trajectory theory of
fermions, with a negative-energy `Dirac sea' and a many-body Dirac equation
[67--70]. Taking for simplicity the free one-body case, we have a Dirac
equation%
\begin{equation}
i\frac{\partial\psi}{\partial t}=-i\boldsymbol{\upalpha}\cdot
\boldsymbol{\nabla}\psi+m\beta\psi\ , \label{Dirac eqn}%
\end{equation}
where $\psi$ is a four-component wave function and $\boldsymbol{\upalpha}$,
$\beta$ are appropriate $4\times4$ matrices. This implies a continuity
equation%
\begin{equation}
\frac{\partial(\psi^{\dag}\psi)}{\partial t}+\boldsymbol{\nabla}\cdot
(\psi^{\dag}\boldsymbol{\upalpha}\psi)=0
\end{equation}
and a de Broglie guidance equation%
\begin{equation}
\frac{d\mathbf{x}}{dt}=\frac{\psi^{\dag}\boldsymbol{\upalpha}\psi}{\psi^{\dag
}\psi}\ . \label{deB_Dirac}%
\end{equation}
A general density $\rho$ evolves by%
\begin{equation}
\frac{\partial\rho}{\partial t}+\boldsymbol{\nabla}\cdot(\rho\mathbf{v})=0\ ,
\end{equation}
where $\mathbf{v}=\psi^{\dag}\boldsymbol{\upalpha}\psi/\psi^{\dag}\psi$, and
we have a quantum equilibrium distribution $\rho=\psi^{\dag}\psi
$.\footnote{Numerical simulations show that quantum relaxation $\rho
\rightarrow\psi^{\dag}\psi$ takes place (on a coarse-grained level) along
similar lines to the low-energy theory \cite{C12}.} If we assume that
$\psi\neq0$, elementary manipulations show that (ref. \cite{Holl93}, p. 505)%
\begin{equation}
\left\vert \mathbf{v}\right\vert \leq1 \label{bound}%
\end{equation}
and the de Broglie velocity is bounded by the speed of light. If instead
$\psi=0$ -- that is, if the four components vanish simultaneously for some
$\mathbf{x}$, $t$ -- then (\ref{deB_Dirac}) is strictly speaking undefined
since both the numerator and the denominator vanish. For example, we might
have an initial wave function $\psi(\mathbf{x},0)$ where the four components
happen to equal the same function $f(\mathbf{x})$ where $f$ has one or more
nodal lines. Thus in principle nodes can occur, where (\ref{deB_Dirac}) is
undefined. However, the bound (\ref{bound}) applies arbitrarily closely to
such points. There is no sign of any divergence close to nodes, and we may
reasonably define (\ref{deB_Dirac}) at nodes by a simple limiting process from
neighbouring points where $\psi\neq0$. For high-energy Dirac particles, then,
the de Broglie velocity is well-behaved close to nodes and our argument for
new physics does not apply.

Even so, our problem cannot be dismissed as an artifact of the low-energy
theory, since diverging velocities still arise in high-energy bosonic field
theory where the de Broglie velocity is again given by a phase gradient -- as
we saw in equations (\ref{deB_Fourier}) and (\ref{deB_QG}) for a scalar field.
Referring to our example of an unentangled field mode $\mathbf{k}$ (Section
2.1), now taken to be on Minkowski spacetime (so that $a=1$), the wave
function $\psi_{\mathbf{k}}(q_{\mathbf{k}1},q_{\mathbf{k}2},t)$ satisfies
\cite{AV07}%
\begin{equation}
i\frac{\partial\psi_{\mathbf{k}}}{\partial t}=-\frac{1}{2}\left(
\frac{\partial^{2}}{\partial q_{\mathbf{k}1}^{2}}+\frac{\partial^{2}}{\partial
q_{\mathbf{k}2}^{2}}\right)  \psi_{\mathbf{k}}+\frac{1}{2}k^{2}\left(
q_{\mathbf{k}1}^{2}+q_{\mathbf{k}2}^{2}\right)  \psi_{\mathbf{k}}\ ,
\end{equation}
with de Broglie velocities%
\begin{equation}
\dot{q}_{\mathbf{k}1}=\frac{\partial s_{\mathbf{k}}}{\partial q_{\mathbf{k}1}%
},\ \ \ \ \dot{q}_{\mathbf{k}2}=\frac{\partial s_{\mathbf{k}}}{\partial
q_{\mathbf{k}2}}\ .
\end{equation}
These match the low-energy equations for a two-dimensional oscillator with
mass $m=1$ and angular frequency $\omega=k$. As in the low-energy particle
theory, the velocities $\dot{q}_{\mathbf{k}1}$, $\dot{q}_{\mathbf{k}2}$
generally diverge close to nodes ($\psi_{\mathbf{k}}=0$). This means that, at
certain points in the field configuration space, the field velocity
$\partial\phi/\partial t$ diverges, which seems physically unacceptable --
just as a diverging particle velocity at certain points in 3-space seems
unacceptable. Similar conclusions apply to the electromagnetic field, whose de
Broglie guidance equation takes the form (in the temporal gauge
\cite{AV24,AV92,AV96})%
\begin{equation}
\frac{\partial\mathbf{A}}{\partial t}=\frac{\delta S}{\delta\mathbf{A}}\ .
\end{equation}
The phase gradient also appears in the guidance equations for other
(non-Abelian) gauge fields \cite{AV24}. We seem obliged to accept that the
pilot-wave theory of high-energy physics breaks down in certain regions of
configuration space, where mathematical regularisation is needed, and close to
which some sort of new physics must set in.

According to our argument, then, pilot-wave dynamics predicts its own demise
at phase singularities where $\psi=0$. We must encounter new physics as we
approach such points. What that new physics might be is a matter for future
research. Pending the development of an improved theory, however, we can
explore simple regularisations of pilot-wave theory and attempt some
confrontation with experiment.

\subsection{Corrections to the Born rule at short distances}

A regularisation of pilot-wave theory was once briefly remarked on (in a
footnote) by Bell (ref. \cite{Bell87}, p. 138). For a general system with wave
function $\psi(q,t)$, we can regularise the de Broglie velocity $v=j/|\psi
|^{2}$ by smearing $j$ and $|\psi|^{2}$ with a narrowly-peaked function so
that the denominator never vanishes. In this simple model, the smeared
$|\psi|^{2}$ becomes the new equilibrium distribution.

Let us define the regularised quantities \cite{AV14}%
\begin{align}
j(q,t)_{\mathrm{reg}}  &  =\int dq^{\,\prime}\ \mu(q^{\,\prime}%
-q)j(q^{\,\prime},t)\ ,\label{j_reg}\\
\left(  |\psi(q,t)|^{2}\right)  _{\mathrm{reg}}  &  =\int dq^{\,\prime}%
\ \mu(q^{\,\prime}-q)|\psi(q^{\,\prime},t)|{^{2}}\ , \label{psi2_reg}%
\end{align}
where $\mu(q^{\,\prime}-q)$ is positive and narrowly-peaked around
$q^{\,\prime}-q=0$ (for example a Gaussian) and $\int dq^{\,\prime}%
\ \mu(q^{\,\prime}-q)=1$. We may then define a regularised de Broglie velocity%
\begin{equation}
v(q,t)_{\mathrm{reg}}=\frac{j(q,t)_{\mathrm{reg}}}{(|\psi(q,t)|^{2}%
)_{\mathrm{reg}}}\ . \label{deB_reg}%
\end{equation}
The usual theory is recovered when $\mu(q^{\,\prime}-q)\rightarrow
\delta(q^{\,\prime}-q)$.

For an ensemble with the same wave function $\psi$, an arbitrary distribution
$\rho(q,t)$ then satisfies%
\begin{equation}
\frac{\partial\rho}{\partial t}+\partial_{q}\cdot(\rho v_{\mathrm{reg}})=0\ .
\label{cont_rho_reg}%
\end{equation}
Now $\psi$ still satisfies the Schr\"{o}dinger equation, hence $\left\vert
\psi\right\vert ^{2}$ still satisfies (\ref{cont_q0}), from which it is
readily shown that%
\begin{equation}
\frac{\partial(|\psi|^{2})_{\mathrm{reg}}}{\partial t}+\partial_{q}%
\cdot\left(  (|\psi|^{2})_{\mathrm{reg}}v_{\mathrm{reg}}\right)  =0\ .
\label{cont_psi2_reg}%
\end{equation}
This takes the same form as (\ref{cont_rho_reg}). Thus, if $\rho=(|\psi
|^{2})_{\mathrm{reg}}$ initially, we have $\rho=(|\psi|^{2})_{\mathrm{reg}}$
at later times. We then have a modified equilibrium distribution with a
smeared Born rule,%
\begin{equation}
\rho(q,t)=(|\psi(q,t)|^{2})_{\mathrm{reg}}\ , \label{Born_reg}%
\end{equation}
where $(|\psi|^{2})_{\mathrm{reg}}$ is positive. If instead $\rho\neq
(|\psi|^{2})_{\mathrm{reg}}$ initially, in appropriate circumstances we can
expect relaxation $\rho\rightarrow(|\psi|^{2})_{\mathrm{reg}}$ (on a
coarse-grained level) as quantified by a decrease of%
\begin{equation}
\bar{H}_{\mathrm{reg}}(t)=\int dq\ \bar{\rho}\ln(\bar{\rho}/\overline
{(|\psi|^{2})_{\mathrm{reg}}})
\end{equation}
(cf. Section 2.1).

Such a regularised theory is of course not intended to be fundamental, but
merely an effective description of some as-yet-unknown physics. This model
predicts a simple smearing of the usual Born distribution at short distances
in configuration space (set by the width of $\mu$). In particular, the new
equilibrium density (\ref{Born_reg}) does not vanish at nodes of $\psi$, in
contrast with the usual prediction $\rho=|\psi|^{2}$ of quantum mechanics. We
may then suggest searching for this effect experimentally.\footnote{If such an
effect were found, it would not be `nonequilibrium' but rather a modification
of equilibrium.}

We can calculate the lowest order correction to the Born rule, in the limit of
a very narrow regularising function $\mu$. Consider smearing a general
function $f(q)$, on $n$-dimensional configuration space, with a narrow
normalised measure%
\begin{equation}
\mu(q^{\,\prime}-q)=\delta_{a}^{n}(q^{\,\prime}-q)=\prod\limits_{i=1}%
^{n}\delta_{a}(q_{\,i}^{\,\prime}-q_{\,i})
\end{equation}
of width $a$ with respect to each degree of freedom $q_{i}$. Writing%
\begin{equation}
I_{a}(q)=\int d^{n}q^{\,\prime}\ \delta_{a}^{n}(q^{\,\prime}-q)f(q^{\,\prime
})=\int d^{n}s\ \delta_{a}^{n}(s)f(q+s)\ ,
\end{equation}
we can expand%
\begin{equation}
f(q+s)=f(q)+s_{i}\partial_{i}f(q)+\frac{1}{2}s_{i}s_{j}\partial_{i}%
\partial_{j}f(q)+...\ .
\end{equation}
Taking%
\begin{equation}
\int d^{n}s\ \delta_{a}^{n}(s)=1\ ,\ \int d^{n}s\ \delta_{a}^{n}%
(s)s_{i}=0\ ,\ \int d^{n}s\ \delta_{a}^{n}(s)s_{i}s_{j}=\delta_{ij}a^{2}\ ,
\end{equation}
we find%
\begin{equation}
I_{a}(q)=f(q)+\frac{1}{2}a^{2}\nabla^{2}f(q)+...\ ,
\end{equation}
where $\nabla^{2}=\sum_{i=1}^{n}\partial_{i}^{2}$. For $f(q)=\left\vert
\psi(q,t)\right\vert ^{2}$ (at fixed $t$) we then have the general result%
\begin{equation}
\left(  \left\vert \psi\right\vert ^{2}\right)  _{\mathrm{reg}}=\left\vert
\psi\right\vert ^{2}+\frac{1}{2}a^{2}\nabla^{2}(\left\vert \psi\right\vert
^{2})+...\ .
\end{equation}
The lowest order correction to the Born rule is proportional to the Laplacian
of $\left\vert \psi\right\vert ^{2}$. We might hope to set experimental bounds
on the (presumably very small) parameter $a$.

As a simple example, consider a one-dimensional oscillator in the first
excited state%
\begin{equation}
\psi_{1}(x)=\left(  \frac{4}{\pi}\right)  ^{1/4}\left(  m\omega\right)
^{3/4}x\exp\left(  -\frac{m\omega x^{2}}{2}\right)  \ .
\end{equation}
The node at $x=0$ is replaced by a minimum%
\begin{equation}
\min\left(  \left\vert \psi\right\vert ^{2}\right)  _{\mathrm{reg}}=a^{2}%
\frac{2}{\pi^{1/2}}\left(  m\omega\right)  ^{3/2}+...\ ,
\end{equation}
with a specific frequency and mass dependence (to lowest order in $a$).
Similarly, for the hydrogen-like state%
\begin{equation}
\psi_{211}(r,\theta,\phi)=-\frac{1}{8\sqrt{\pi a_{0}^{3}}}\frac{r}{a_{0}%
}e^{-r/2a_{0}}(\sin\theta)e^{i\phi}\ ,
\end{equation}
on the nodal line ($z$-axis) we find%
\begin{equation}
\left(  \left\vert \psi_{211}\right\vert ^{2}\right)  _{\mathrm{reg}}%
=a^{2}\frac{1}{32\pi a_{0}^{5}}e^{-r/a_{0}}+...\ ,
\end{equation}
with a specific dependence on $r$. The second example is unrealistic as the
physics is superseded by the high-energy Dirac theory. However, the first
example does apply realistically to high-energy bosonic field theory, where as
noted a Fourier mode $\mathbf{k}$ of a massless scalar field (on Minkowski
spacetime) corresponds mathematically to a two-dimensional oscillator with
$m=1$ and $\omega=k$. In principle, such effects could be tested by careful
measurements of the electromagnetic field (perhaps in optical cavities).

\subsection{Instability of quantum equilibrium at short times}

We may also briefly discuss a natural generalisation of the above model to a
time-dependent regularising function $\mu(q^{\,\prime}-q,t)$ \cite{AV14}. If
there is new physics at nodes of $\psi$, it seems plausible that the function
$\mu$ could be time dependent -- for example, during high-energy collisions
taking place over very short times.

Let us again define a regularised de Broglie velocity (\ref{deB_reg}), with
$j_{\mathrm{reg}}$ and $(|\psi|^{2})_{\mathrm{reg}}$ as in (\ref{j_reg}) and
(\ref{psi2_reg}), but with $\mu(q^{\,\prime}-q)$ replaced by $\mu(q^{\,\prime
}-q,t)$. We now find that, instead of (\ref{cont_psi2_reg}), $(|\psi
|^{2})_{\mathrm{reg}}$ satisfies%
\begin{equation}
\frac{\partial(|\psi|^{2})_{\mathrm{reg}}}{\partial t}+\partial_{q}%
\cdot\left(  (|\psi|^{2})_{\mathrm{reg}}v_{\mathrm{reg}}\right)  =s\ ,
\label{cont_psi2_reg_s}%
\end{equation}
with a `source'%
\begin{equation}
s(q,t)=\int dq^{\prime}\ \frac{\partial\mu(q^{\,\prime}-q,t)}{\partial t}%
|\psi(q^{\,\prime},t)|^{2}\ .
\end{equation}
There is now a mismatch between the continuity equation (\ref{cont_rho_reg})
for $\rho$ (which of course still holds) and the continuity equation
(\ref{cont_psi2_reg_s}) for $(|\psi|^{2})_{\mathrm{reg}}$. As in semiclassical
gravity with small quantum-gravitational corrections (Section 3.4), an initial
equilibrium distribution $\rho=(|\psi|^{2})_{\mathrm{reg}}$ can evolve into a
final nonequilibrium distribution $\rho\neq(|\psi|^{2})_{\mathrm{reg}}$. In
the extended regularised theory, with a time-dependent smearing function, the
Born rule is unstable.

The extended model suggests that quantum nonequilibrium could be created in a
high-energy collider experiment. If a collision occurs during an approximate
time interval $(t_{i},t_{f})$, we might take $\mu$ to be static ($\partial
\mu/\partial t=0$) at times $t<t_{i}$ and $t>t_{f}$, so that we have an
equilibrium distribution $(|\psi|^{2})_{\mathrm{reg}}$ before and after the
collision (where $(|\psi|^{2})_{\mathrm{reg}}\simeq|\psi|^{2}$ if $\mu$ has
negligible width). If $\partial\mu/\partial t\neq0$ during $(t_{i},t_{f})$, an
incoming equilibrium state $\rho_{\mathrm{in}}=(|\psi|^{2})_{\mathrm{reg}}$
can evolve into an outgoing nonequilibrium state $\rho_{\mathrm{out}}%
\neq(|\psi|^{2})_{\mathrm{reg}}$.\footnote{For a worked example in a
cosmological context see ref. \cite{AV14}.}

\subsection{Testing the Born rule in high-energy collisions}

We have seen that the need for regularisation does not apply directly to
trajectories of Dirac fermions. However, if these are coupled to regularised
bosonic fields, where the latter have a modified or unstable Born rule, we can
expect all particle species to display corrections to the Born rule. Such
corrections could of course manifest in many ways. How might such corrections
be most conveniently observed? We have two suggestions. The first involves
testing for smeared zeros of differential scattering cross sections. The
second involves testing for anomalous spin or polarisation probabilities. We
briefly review these.

In high-energy physics we typically calculate S-matrix elements from initial
states $\left\vert i\right\rangle $ to final states $\left\vert f\right\rangle
$. These take the schematic form (omitting overall normalisation factors)%
\begin{equation}
\left\langle f\right\vert \hat{S}\left\vert i\right\rangle \sim\delta
^{4}(p_{f}-p_{i})\mathcal{M\ },
\end{equation}
where $\mathcal{M}$ is the Feynman amplitude (obtained from Feynman rules) and
$\delta^{4}(p_{f}-p_{i})$ is an energy-momentum conserving delta-function.
Applying the usual Born rule, we may then derive differential cross sections
of the quantum-theoretical form%
\begin{equation}
\left(  \frac{d\sigma}{d\Omega}\right)  _{\mathrm{QT}}\sim\left\vert
\mathcal{M}\right\vert ^{2}\ .
\end{equation}
Now, in a de Broglie-Bohm formulation, the outcome of a high-energy collision
experiment at a final time $t_{f}$ is determined by the initial conditions
$\psi(q,t_{i})$, $q(t_{i})$ at time $t_{i}$, where the configuration $q$
includes whatever particles and fields may be relevant.\footnote{These may
include degrees of freedom in the measuring apparatus.} For an ensemble of
experiments with a given $\psi(q,t_{i})$, the distribution of outcomes is
determined by the initial distribution $\rho(q,t_{i})$ of configurations.
Thus, for a given initial quantum state, $d\sigma/d\Omega$ is determined by
$\rho(q,t_{i})$. Should $\rho(q,t_{i})$ be unequal to the usual Born
distribution $\left\vert \psi(q,t_{i})\right\vert ^{2}$, we can expect to find
a departure%
\begin{equation}
\frac{d\sigma}{d\Omega}\neq\left(  \frac{d\sigma}{d\Omega}\right)
_{\mathrm{QT}} \label{dsigma_noneq}%
\end{equation}
from the usual differential cross section. In particular, we may expect to
find a smearing of the angular dependence and (perhaps) an erasure of zeros.
Experimentally, then, we suggest carefully probing the angular distribution of
outcomes close to points where $\left\vert \mathcal{M}\right\vert ^{2}$
vanishes, in order to discover if the distribution truly goes to zero -- or if
instead it is smeared out at small angular scales.

Should anomalies in cross sections be observed, it will be natural to ask if
they are really caused by corrections to the Born rule or if instead they are
caused by some unexpected correction to the Hamiltonian. For this reason, spin
or polarisation probabilities provide a cleaner test, as follows.

Any two-state quantum system has observables $\hat{\sigma}\equiv
\mathbf{m}\cdot\boldsymbol{\hat{\upsigma}}$ with eigenvalues $\sigma=\pm1$,
where $\mathbf{m}$ is a unit vector on the Bloch sphere and $\boldsymbol{\hat
{\upsigma}}$ is a Pauli spin operator. According to quantum theory, if
$\hat{\sigma}$ is measured over an ensemble with density operator $\hat{\rho}%
$, we will find an expectation value%
\begin{equation}
E_{\mathrm{QT}}(\mathbf{m})\equiv\left\langle \mathbf{m}\cdot\boldsymbol{\hat
{\upsigma}}\right\rangle =\mathrm{Tr}\left[  \hat{\rho}\left(  \mathbf{m}%
\cdot\boldsymbol{\hat{\upsigma}}\right)  \right]  =\mathbf{m}\cdot
\mathbf{P\ ,} \label{E_QT}%
\end{equation}
where $\mathbf{P}=\langle\boldsymbol{\hat{\upsigma}}\rangle=\mathrm{Tr}\left[
\hat{\rho}\boldsymbol{\hat{\upsigma}}\right]  $ is the average polarisation.
The outcome $\sigma=+1$ then has a Born-rule probability%
\begin{equation}
p_{\mathrm{QT}}^{+}(\mathbf{m})=\frac{1}{2}\left(  1+E_{\mathrm{QT}%
}(\mathbf{m})\right)  =\frac{1}{2}\left(  1+P\cos\theta\right)  \ ,
\label{Malus_gen}%
\end{equation}
where $\theta$ is the angle between $\mathbf{m}$ and $\mathbf{P}$. For
spin-1/2 particles, values $\sigma=\pm1$ represent spin (in units $\hbar/2$)
up or down a spatial axis $\mathbf{m}$, while for photons they represent
polarisation parallel or perpendicular to a spatial axis $\mathbf{M}$ (with
$\theta$ corresponding to an angle $\Theta=\theta/2$ in 3-space). If the Born
rule is violated, we can expect to observe deviations from the sinusoidal
modulation (\ref{Malus_gen}).

For example, if single photons pass through a linear polariser, the emerging
beam will be fully polarised ($P=1$). If the beam then impinges on a second
linear polariser, at angle $\Theta$ with respect to the first,
(\ref{Malus_gen}) predicts a transmission probability%
\begin{equation}
p_{\mathrm{QT}}^{+}(\Theta)=\cos^{2}\Theta\,. \label{cos2}%
\end{equation}
If we vary $\Theta$ over an ensemble of measurements, deviations from
(\ref{cos2}) would signal an unambigous violation of the Born rule.

It should be emphasised that the linearity in $\mathbf{m}$ of the quantum
expectation value (\ref{E_QT}) is equivalent to additive expectation values
for incompatible or non-commuting observables (such as spins along different
axes) \cite{AV04a}. The latter is a deep and remarkable property of quantum
states \cite{Bell66}, which is violated in quantum nonequilibrium (for general
deterministic hidden-variables theories) \cite{AV04a}. A nonlinear expectation
value%
\begin{equation}
E(\mathbf{m})=\epsilon+P_{i}m_{i}+Q_{ij}m_{i}m_{j}+R_{ijk}m_{i}m_{j}%
m_{k}+\ ... \label{nonlin}%
\end{equation}
(summing over repeated indices, where in Bloch space $\epsilon$ is a constant
scalar, $P_{i}$ is a constant vector, and $Q_{ij}$, $R_{ijk}$, ... are
constant tensors) then provides a simple and universal signature of quantum
nonequilibrium \cite{AV07}, with an anomalous outcome probability%
\begin{equation}
p^{+}(\mathbf{m})=\frac{1}{2}\left(  1+E(\mathbf{m})\right)  \ .
\end{equation}
Note that the vector $P_{i}$ in (\ref{nonlin}) can differ from the
quantum-theoretical polarisation vector $\mathbf{P}^{\mathrm{QT}}$ expected
from the usual quantum state preparation.

For a two-state quantum system deviations from the Born rule can also be
parameterised by a spherical harmonic expansion%
\begin{equation}
p^{+}(\theta,\phi)=\sum_{l=0}^{\infty}\sum_{m=-l}^{+l}b_{lm}Y_{lm}(\theta
,\phi)\ ,
\end{equation}
where $\mathbf{m}$ is specified by angular coordinates $(\theta,\phi)$ on the
Bloch sphere and we may take $\mathbf{P}^{\mathrm{QT}}$ to point along
$+z$.\footnote{The reality of $p^{+}$ requires $b_{lm}^{\ast}=(-1)^{m}%
b_{l(-m)}$.} In quantum equilibrium $p^{+}=p_{\mathrm{QT}}^{+}$ and the only
non-vanishing coefficients are%
\begin{equation}
b_{00}^{\mathrm{QT}}=\sqrt{\pi}\ ,\ b_{10}^{\mathrm{QT}}=P\sqrt{\pi/3}\ .
\end{equation}
Anomalous coefficients can in principle be calculated from a hidden-variables
theory (such as pilot-wave theory) with corrections to the Born rule.

Experimentally, we would hope to set upper bounds on the magnitudes of the
parameters $\epsilon$, $Q_{ij}$, $R_{ijk}$, ... , or equivalently, upper
bounds on the magnitudes of the coefficients $b_{lm}$ ($lm\neq00,10$). This
could involve careful analysis of spin or polarisation data from high-energy collisions.

In recent years, high-energy experiments have been repurposed as novel tests
of entanglement and Bell's inequality \cite{Barr24}. Such experiments may also
be repurposed as tests of the Born rule at high energies and at short times.
Strong interactions take place over timescales of order $10^{-23}\,\mathrm{s}%
$. To probe the Born rule at such short times, we might consider photons
emerging from a strong-interaction process (where photons can be radiated by
quarks). By monitoring their polarisation probabilities, we may search for
nonlinear expectation values (\ref{nonlin}) and associated violations of the
Born rule.

\section{Conclusion}

In the spirit of our opening quote from Feynman's 1964 lectures on \textit{The
Character of Physical Law} \cite{Feyn67}, we have seen that de Broglie's
pilot-wave theory is richly suggestive of new physics. We have summarised how
the theory can naturally depart from quantum mechanics in three specific
domains: the early universe, quantum gravity, and high-energy physics. In all
three cases, we are able to predict new effects that are potentially
observable or at least amenable to constraint by experiment.

Our quote from Feynman was taken from the final chapter of those lectures,
entitled `Seeking New Laws', which attempts to outline the various ways in
which new laws (hence new physics) are most likely to be found. According to
Feynman, it is often underestimated just how difficult it is to find new laws
that account for what is known while at the same time making new predictions:

\begin{quotation}
`If you can find any other view of the world which agrees over the entire
range where things have already been observed, but disagrees somewhere else,
you have made a great discovery. It is very nearly impossible, but not quite,
to find any theory which agrees with experiments over the entire range in
which all theories have been checked, and yet gives different consequences in
some other range ... .' (ref. \cite{Feyn67}, pp. 171--2)
\end{quotation}

The often haphazard and serendipitous history of science suggests that we be
wary of making hard and fast rules about how science should proceed. Nor
should we be overly confident about which tactics are most likely to succeed.
However, it seems fair to suggest that the de Broglie-Bohm or pilot-wave
approach to quantum physics deserves more attention than it has generally
received, in particular as regards the search for new physics. As Feynman
rightly emphasised, to have a theory that agrees with experiment in domains
thus-far explored, and which naturally suggests new physics in domains
yet-to-be explored, is something of a rarity -- a valuable asset whose
implications deserve to be taken seriously. Whether or not that suggested new
physics will ever be observed is, of course, a matter for further research and experiment.

\end{document}